# Variable-Path-Length FTIR of *E. coli* in Aqueous Media

Jonathan Matsuura[1,2], Andrew Huang[3], Jaehyeon Kim[1,4], Ching-Ping Chang[1,4], Kai Zhang[3], Yingjie Zhang[1,4,5]*

[1] Materials Research Laboratory, University of Illinois, Urbana, Illinois 61801, United States

[2] Department of Chemistry, University of Illinois, Urbana, Illinois 61801, United States

[3] Department of Biochemistry, University of Illinois, Urbana, Illinois 61801, United States

[4] Department of Materials Science and Engineering, University of Illinois, Urbana, Illinois 61801, United States

[5] Beckman Institute for Advanced Science and Technology, University of Illinois, Urbana, Illinois 61801, United States

*Correspondence to: yjz@illinois.edu

**ABSTRACT:** Transmission infrared spectroscopy has been widely used for chemical analysis of biological samples in aqueous environments. However, its scope of applications has been limited by the path length, which is either too large or fixed, posing challenges for analyzing highly absorbing or heterogeneous samples. In this work, a mid-infrared (mid-IR) optical fiber probe was used for Fourier transform infrared (FTIR) micro-spectroscopy of aqueous *Escherichia coli* (*E. coli*) samples, providing continuous tuning of optical path length and sampling of near-surface and bulk regions. The mid-IR absorbance of the protein signal at 1548 $cm^{-1}$ increased linearly with path length, consistent with the Beer-Lambert law. Path-length dependent spectra were used to calculate the spatially heterogeneous absorption coefficient of *E. coli* suspensions in aqueous media. The results demonstrate the ability of our fiber-based technique to resolve signals originating from different depths into the biological solution.

## TOC Graphic

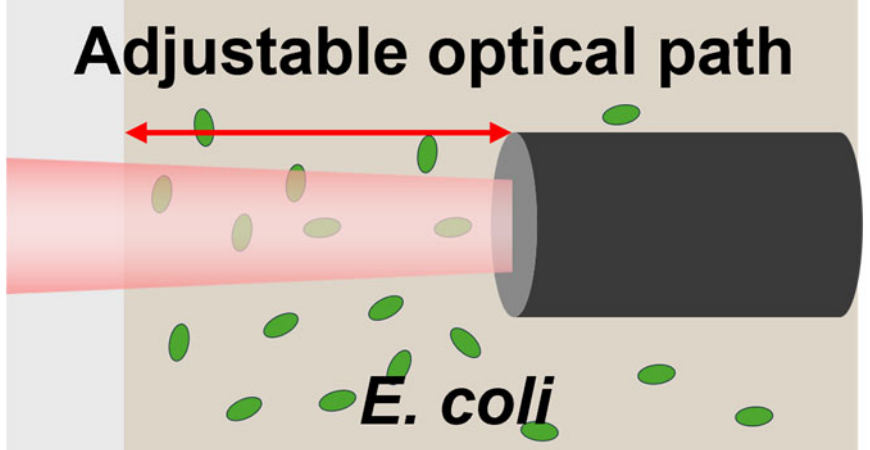

## INTRODUCTION

FTIR is a vibrational spectroscopy technique, effective at identifying organic functional groups.[1–3] The molecular vibrations of many functional groups fall into mid-IR (2.5 to 20 μm), making FTIR suitable at chemical fingerprinting. Application of FTIR spectroscopy to biological samples has revealed the myriad of polar molecules that make up organisms.[4–7] However, due to the strong mid-infrared absorption of water, many existing FTIR measurements of biological samples have been limited to the dried state. While such measurements produced some chemical information, the drying process can significantly perturb the intrinsic structure of biological samples which containing water as the largest constituent.[8] Therefore, there is interest in developing FTIR measurements for biological systems in native aqueous environments.[9]

To chemically detect live cells in aqueous environments, several approaches have been employed. attenuated total reflectance (ATR)-FTIR relies on an evanescent wave from an internal reflection element (IRE) to produce the signal, resulting in path lengths on the order of 1 to 2 μm.[10,11] While this can avoid the complete absorption of signal by water in mid-IR, it is limited to measuring surface regions. For cells with a larger diameter than this path length, the measured signal only captures partial information of the cell. Further, the small path length leaves the ATR-FTIR prone to unintentional surface-induced structural and compositional changes of the biological systems. Atomic force microscopy-IR spectroscopy can analyze biological samples with subcellular resolution but are generally limited to the very surface of the sample,[12,13] making it difficult, if at all possible, to examine the internal structure of whole, live cells. A microfluidic cell (MFC), in which a thin liquid layer is confined between two IR-transparent windows, enables measurements of live cells and their metabolic activity in aqueous environments. However, the path length of the MFC is typically fixed and chosen to be sufficiently small (<10 μm) to avoid the overwhelming absorbance from water.[14,15] Cells larger than the path length cannot be introduced to the microfluidic channel, while cells with a size slightly smaller than the path length can experience compression and perturbation.[16–18] The difficulty in precisely controlling the path length further compromises the reproducibility and quantification capability.[19] Confocal Raman spectroscopy has also been widely applied,[20,21] offering micrometer-level spatial resolution due to the shorter wavelengths of the excitation laser.[22] However, its inherently weak scattering cross section and susceptibility to fluorescence background limit its detectivity.[23,24]

In this study, seeking to measure live cells in aqueous environments, we fabricated a mid-IR fiber microprobe for insertion into the cell solution, enabling continuous, precise control of the local path length. Our prior work benchmarked this fiber-based endoscopic FTIR (endo-FTIR) method by measuring pure water at a series of path lengths.[25] While *E. coli* (1–2 μm in diameter) is small enough to be (partially) sensed by ATR-FTIR,[26,27] such studies would only be able to detect *E. coli* that is settled on the IRE, lacking characterization of cells that are dispersed in the bulk. Here we study *E. coli* cells in aqueous solution, which serves as model prokaryotic cells to demonstrate the capability of our endo-FTIR approach in measuring biological systems. Unlike an MFC, the liquid sample was unconfined outside of the local, micron-scale optical path in our endo-FTIR method, allowing for reversible control of the path length by the stage translation *in situ*. By moving the position of the microbial sample while the mid-IR fiber microprobe was immersed in the solution, the FTIR spectra of *E. coli* HST08 suspensions were recorded at a series of path lengths, enabling depth-resolved chemical profiling (Figure 1). The results demonstrated the efficacy of endo-FTIR for measuring suspended microorganisms in aqueous media without the surface-sensitivity restriction of ATR and the confined geometry of thin liquid cells.

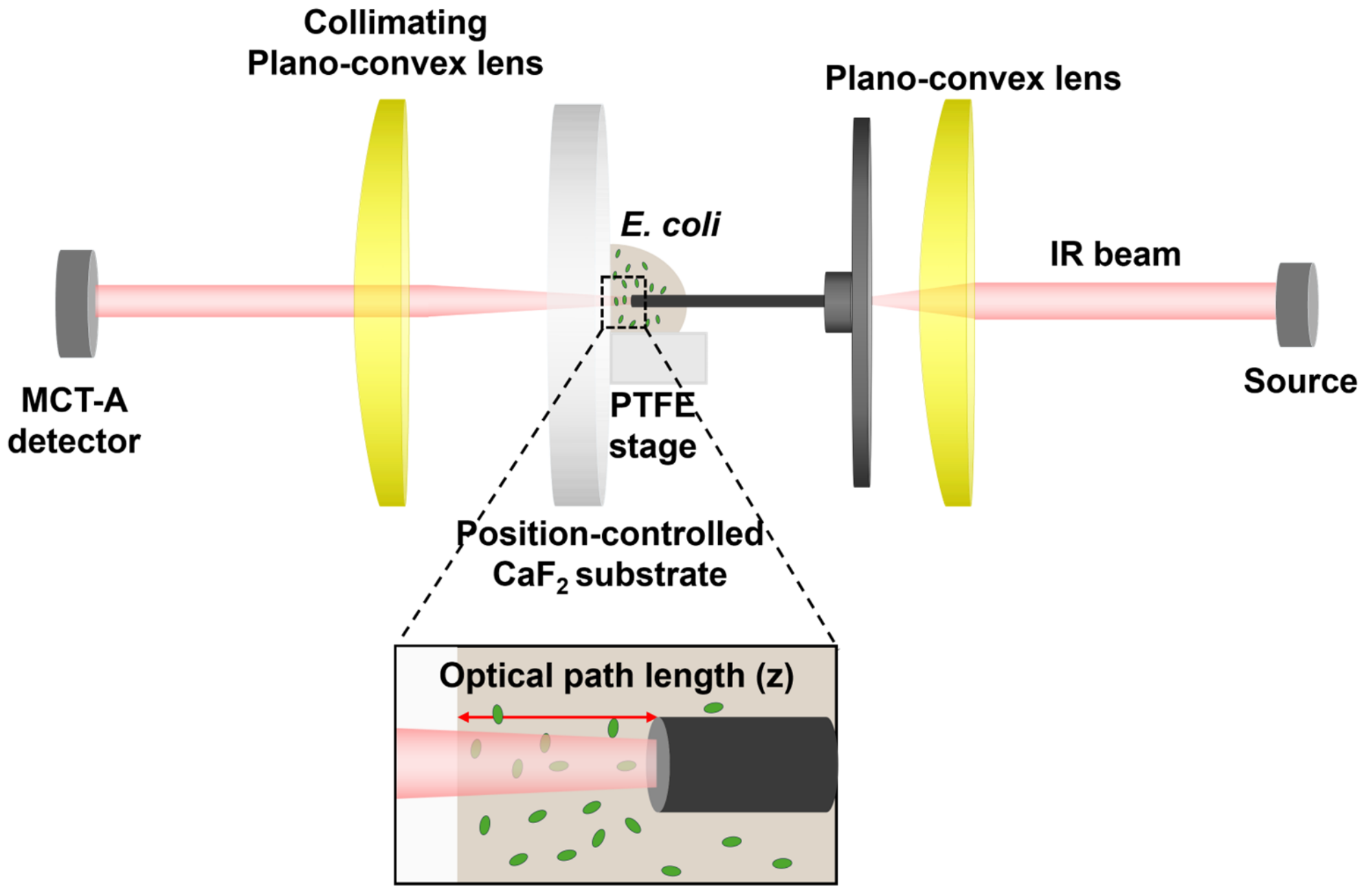

**Figure 1.** Schematic illustration of the endo-FTIR setup.

## METHODS

### Fiber Microprobe Manufacturing

$As_2Se_3$ optical fiber (IRflex Corporation, IRF-Se-100R) fiber was cut to a length of 1.5 cm. The fiber was immersed in acetone (Fisher Chemical) for 10 minutes to soften the polyacrylate coating. The coating was removed by nicking with a razor blade and sliding it off the fiber. The fiber was inserted into a 26-gauge needle and secured with epoxy (Devcon 5 Minute Epoxy) to improve rigidity and mechanical stability. Following curing, the input and output ends of the fiber were polished to submicrometer roughness using silicon carbide and alumina lapping sheets (Thorlabs, LF5P, LF3P, LF1P, and LF03P). Through this polishing the roughness of the fiber ends was reduced to an average areal roughness below 0.2 µm (Figure S1), resulting in a significant increase in mid-IR transmission through the fiber microprobe (Figure S2). The plastic hub of the needle was then thinned with a razor blade to allow coupling to the optical setup.

### Optical Setup

The core opto-mechanical components were assembled onto a small optical breadboard (Thorlabs, MB8, 8 × 8 × 1/2 in.) and inserted into a Thermo Nicolet iS50 spectrometer. A ZnSe plano-convex lens (Thorlabs, LA7656-E4, 50.1 mm FL) was used to couple the IR light source to the fiber microprobe. The fiber microprobe was inserted into a ferrule adapter plate (Thorlabs, SM1FCM, made of Al 6061-T6) fixed to an XY translator (Thorlabs, ST1XY-D) for fine alignment control, all attached to a mount with adjustable rotation of the supporting rod. Electrical tape was used to seal possible light leaks. A $CaF_2$ window (Thorlabs, WG51050, diameter: 25.4 mm, thickness: 5.0 mm) was secured into a mirror mount (Thorlabs, FMP1) and then fixed to an XYZ translation stage (Thorlabs, MT3A) to achieve 3D position control with ~1 µm precision. A PTFE block was attached to the window with Kapton tape to serve as a stage for liquid samples. Another ZnSe plano-convex collimating lens (Thorlabs, LA7656-E4, 50.1 mm FL) was used to focus the output to the detector.

**Alignment**

Following mounting of the fiber microprobe, the intensity of the transmitted light was monitored through the spectrometer with Omnic software. The angle of the optical fiber microprobe to the beam and detector was adjusted using the screw mount. The position of the fiber relative to the first plano-convex lens was controlled using the translation stage. The orientation of the collimating plano-convex lens was controlled via its mount. Alignment was adjusted to maximize the intensity of the light transmitted through the fiber. Following the alignment, the $CaF_2$ window sample stage was placed into the optical path between the collimating lens and the fiber microprobe. The efficacy of the coupling was tested by applying carbon paint (Ted Pella) to the tip of the fiber to block transmission. It was found that, upon adding paint, no signal was transmitted to the detector (Figure S2). Carbon paint was removed by rinsing with acetone prior to the following testing.

**Optical Measurements**

A spectrum of the stage in air was taken to serve as a background prior to deposition of sample on the stage. Following addition of aqueous sample, pure water, DPBS (Dulbecco's phosphate buffered saline), or *E. coli* dispersed in DPBS, the intensity at the water combination mode (2127 $cm^{-1}$) was divided by the corresponding intensity of the air background and used to estimate the optical path length with the following equation, from the Beer-Lambert Law.

$$T = e^{-\alpha z}, \tag{1}$$

where T is the transmittance following corrections, $\alpha$ is the absorption coefficient of the sample, and z is the optical path length. While not nominally homogeneous, the bulk section of *E. coli* suspensions demonstrated linearity with optical path, so the Beer-Lambert Law was used to extract an approximation of the absorption coefficient. It was assumed that the beam was normal to the surfaces of the fiber output and the substrate, yielding the following form of the Fresnel equation (see Hecht, Optics, 5th ed., Eq. 4.67).[28]

$$R = \left(\frac{n_1 - n_2}{n_1 + n_2}\right)^2, \tag{2}$$

where $n_1$ and $n_2$ are the refractive indices of the incident and transmitting medium, respectively.

The mean average of the refractive indices over the range of spectroscopy (2.5 to 10 μm) were calculated for the optical media.[29–31] The relevant interfaces are: 1) between the fiber ($n \approx 2.7$) and the sample ($n \approx 1.0003$ for air and 1.32 for water), and 2) between the sample and the $CaF_2$ substrate ($n \approx 1.38$). The reflectance values at these two interfaces are $R_1$ and $R_2$, respectively. The interfacial transmittance $T_{interfacial}$ can be calculated by multiplying the losses due to each reflectance.

$$T_{interfacial} = (1 - R_1) * (1 - R_2). \tag{3}$$

The measured transmittance, $T_{meas}$, is related to the actual transmittance through the aqueous sample, T, via the following formula:

$$T_{meas} = \frac{I_{aqueous}}{I_{air}} = T * \frac{T_{interfacial,aqueous}}{T_{interfacial,air}}, \tag{4}$$

where $I_{aqueous}$ is the measured intensity of the aqueous sample, and $I_{air}$ is the measured intensity of the air spectrum. Using eq 3, we obtained $T_{interfacial,air}$ = 0.769 and $T_{interfacial,aqueous}$ = 0.882. Since $T_{interfacial,aqueous}$ / $T_{interfacial,air}$ ≈ 1.15, eq 4 reveals that $T_{meas}$ is 15% higher than T.

The optical path length through any aqueous sample can be precisely calculated by combining Eqs 1 and 4, using water vibrational modes (Figure S3). For water at 2127 $cm^{-1}$, representing the maximum of the combination mode, the absorption coefficient, α has been reported as 420 $cm^{-1}$.[29] Using this mode, the optical path length can be written as:

$$z = \frac{-\ln(T_{meas}) + \ln(1.15)}{420\ cm^{-1}}. \tag{5}$$

The absorbance at 2127 $cm^{-1}$ was used because it has lower absorption than the OH stretching and the HOH bending modes, allowing for measurement of longer path length (z > 15 μm). It also has minimal overlap with *E. coli* or DPBS. The sample stage was manually adjusted to the desired path length prior to recording the spectrum, using the value from the spectral preview, representing the average of the four most recent scans. Upon any replacement of sample, a new air spectrum was collected as a baseline, and subsequently used to calculate the path length for the sample.

Spectra were collected with a resolution of 4 $cm^{-1}$.

**Spectral Analysis**

Spectra were processed using OriginPro. Spectra were considered from 3000 to 1000 $cm^{-1}$, where the mid-IR transmission signal can be detected without complete absorption by either the fiber or water. Transmittance spectra were produced by dividing the liquid sample spectrum by the corresponding air spectrum to remove signal from carbon dioxide and water vapor from the environment. The resulting spectra were further divided by the transmittance spectra of DPBS at the same calculated path length to remove signal from the aqueous media and exogenous phosphate. An illustration of this process is shown in Figure S4. The transmittance spectra were converted to absorbance of *E. coli* via:

$$A_{E.coli} = -\log\left(\frac{T_{E.coli,DPBS}}{T_{DPBS}}\right). \tag{6}$$

Prior to analysis of the absorbance spectra, baseline subtraction was performed using the Peak Analyzer feature in OriginPro to remove residual water signal due to subtle, undetected changes in alignment over time. Signal corresponding to HOH bending of water at 1650 $cm^{-1}$, with overlapping amide I peak, was excluded, as the dominance of water signal precludes the reliable analysis of *E. coli* signal at this position.

The natural log of the transmittance was plotted vs the path length z. The slope of the resulting linear regression equals to $-\alpha_{bulk}$ (absorption coefficient of the bulk liquid) for each wavenumber of the spectrum, averaged over the number of spectra:

$$\alpha_{bulk} = -d[Ln(T)]/d(z). \tag{7}$$

Through this calculation, dependence on path length can be removed, enabling precise quantification of $\alpha_{bulk}$. Spectra representative of the bulk *E. coli* were produced by subtracting the absorption coefficient of the DPBS solution from that of the *E. coli* suspension in DPBS solution. For comparison, the absorbance spectra of samples at path length of 4.9 μm, used as representation of the near-surface region, were converted to absorption coefficient $\alpha$ using eq 8, a rearrangement of the Beer-Lambert law:

$$\alpha_{near-surf} = \frac{A_{near-surf} * \ln(10)}{z} \tag{8}$$

## *E. coli* Sample Preparation

*E. coli* HST08 were used as the model cells. The competent E. coli cells were transformed with plasmids encoding for ampicillin resistant genes, plated in an ampicillin-containing agar plate and incubated in 37 degrees overnight. Once colonies grew, colonies were inoculated in 5 mL of lysogeny broth (LB) media, supplemented with 5 μL of 1000X ampicillin, equivalent to 100 mg/mL before dilution. After 16–20 hours, samples were removed from incubator and suspended in DPBS to the desired concentration, based on the optical density of the sample at 600 nm, measured with Nanodrop 2000c spectrophotometer (Thermo Fisher).

# RESULTS AND DISCUSSION

Following previous demonstration of the technique with pure water,[25] in this work we conducted endo-FTIR on biological samples in aqueous solution. Maximal path length was selected at the point of saturation of the water HOH bending mode (1639 $cm^{-1}$), ~17 μm, to ensure that signals in the fingerprint region, 1600 to 1000 $cm^{-1}$, will not be overwhelmed by the HOH bending mode. Bovine serum albumin (BSA), a water-soluble protein, was measured at 5 vol.% to test the fiber microprobe's ability to detect protein in aqueous solution. Signals consistent with previously reported spectra were detected.[32] Linear correlation between the absorbance and path length was observed, consistent with the Beer-Lambert law (Figure S5).

A high concentration of cells was chosen to represent concentrated biomass samples and ensure detection through FTIR spectroscopy. Cultured *E. coli* was diluted to a concentration of ~4.5 vol.%, determined via UV-Vis spectroscopy.[33,34] A representative endo-FTIR spectrum of the sample is presented in Figure 2.

Signals present at 2960, 2920, and 2850 $cm^{-1}$, consistent with C–H bonds, were detected in BSA and *E. coli* samples.[35] The amide II band, corresponding to N–H bending of amide bonds, was detected at ~1548 $cm^{-1}$ and used as the diagnostic for detection of *E. coli* or protein features. Bands at 1456 and 1400 $cm^{-1}$ were detected in BSA and *E. coli* samples, corresponding to C–H stretching and $COO^{-}$, respectively. A peak at 1241 $cm^{-1}$ was attributed to random coils of the protein secondary structure.[36] Though sometimes assigned to asymmetric phospholipid stretching,[37] we conclude that the signal is most likely due to amide III as it is present in BSA solution (Figure S5) where phosphorus is absent.

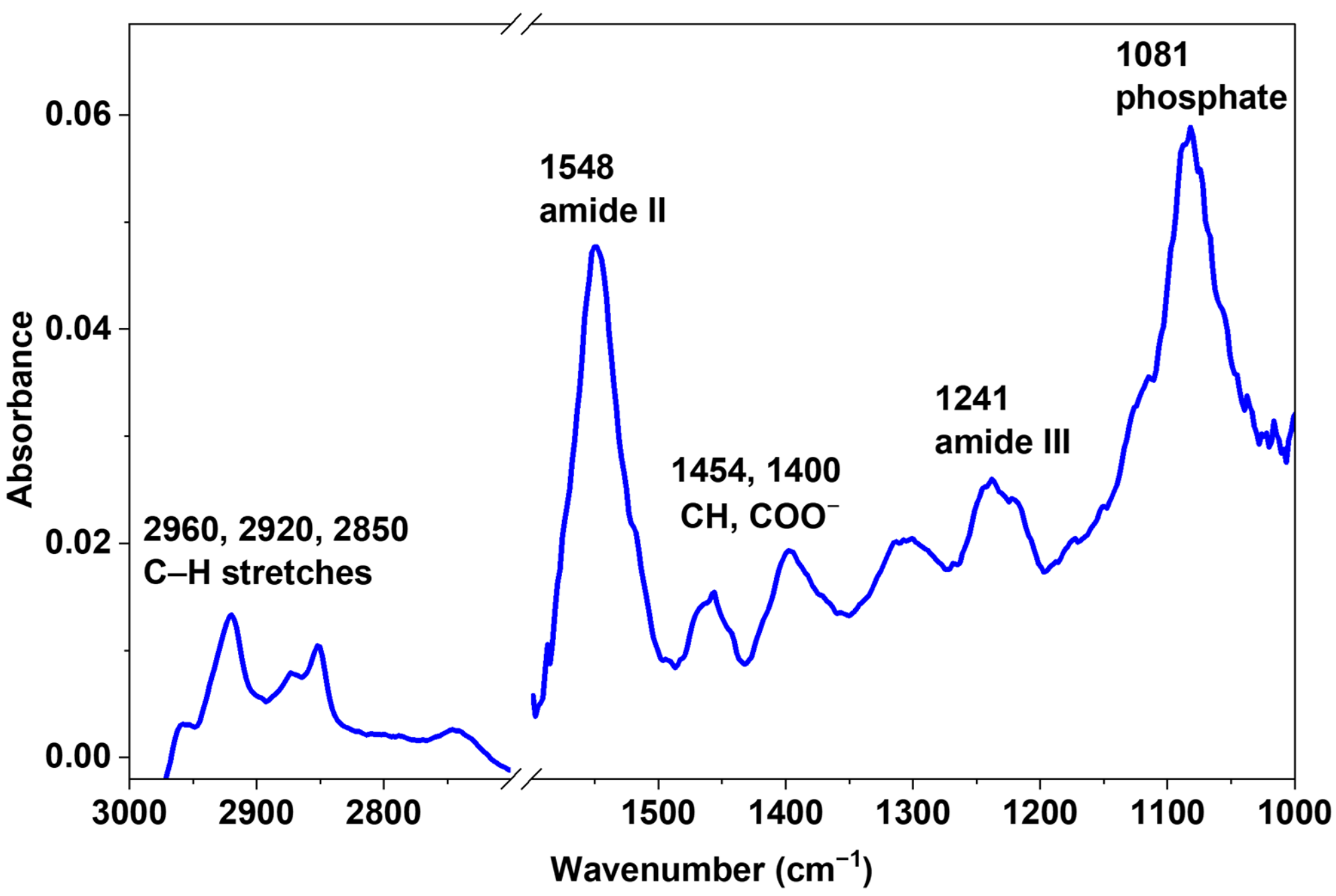

**Figure 2.** DPBS-subtracted IR absorbance spectrum of 4.5 vol.% *E. coli* suspension at z = 17 µm. Peaks corresponding to biological molecules are annotated.

Spectra were taken with 512 scans during benchmarking of the microprobe. The signal to noise ratio of the amide II peak at 1548 $cm^{-1}$ was calculated as 189.9, using the standard deviation of the absorbance in the 2600–2700 $cm^{-1}$ range as the noise. Treating the *E. coli* sample as 4.5 vol.%, the limit of detection (LOD) can be extrapolated from the concentration that would produce a signal to noise ratio of 3, giving an apparent LOD of 0.07 vol.% *E. coli* for detection over the water background. This is in line with a spectrum taken of 8-fold diluted *E. coli* being clearly discernible (Figure S6). Following this benchmarking, spectra were taken with 64 scans to lower the acquisition time from 32 minutes to 4 minutes, avoiding possible E. coli settling on the PFTE stage and improving the accuracy of variable-path-length measurements.

A series of spectra were collected at decreasing path lengths by approaching the substrate to the fiber microprobe (Figure 3). Spectra were measured with a variable number of points to capture different length intervals, consisting of three points (Trial 1), six points (Trial 2), and twelve points (Trial 3). The absorbance is approximately a linear function of the path length, indicating that the cells are nearly uniformly distributed in the bulk solution. The differences in the slopes of the linear regressions suggest that the effective cell concentration of the three trials differed, despite nominally identical preparation. Because the samples were measured as suspensions, partial settling of cells prior to or during measurement likely resulted in differences in the concentration of cells sampled within the optical path.

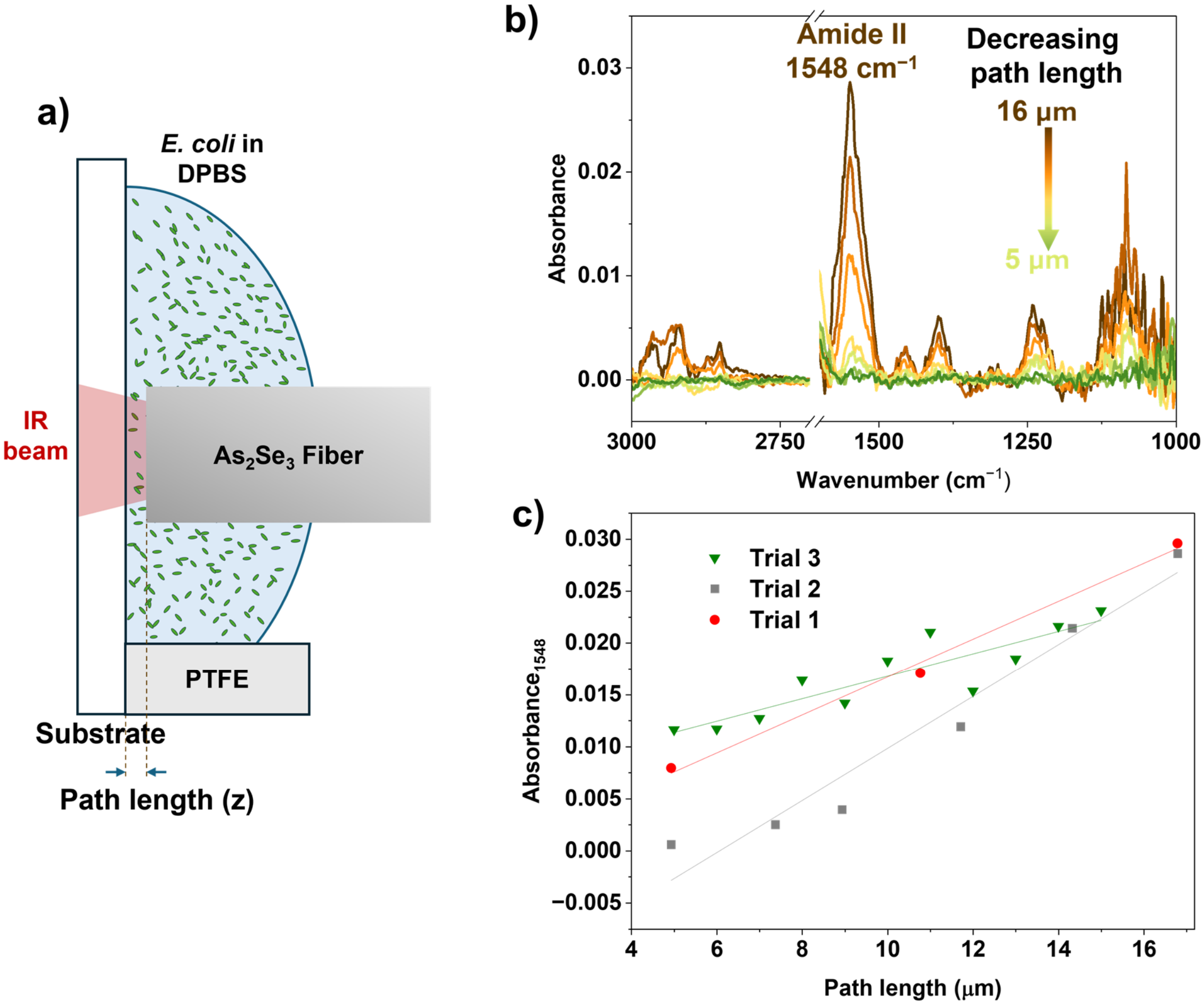

**Figure 3.** Path length dependence of absorbance. (a) Schematic diagram of the variable-path-length endo-FTIR setup. **(**b) Representative spectrum of 4.5 vol.% *E. coli* suspension at decreasing path length. c) Absorbance at 1548 cm$^{-1}$ vs path length from individual measurements, with linear regression.

Through linear regression of the absorbance value vs the optical path length at z ≥ 4.9 μm, we determined the absorption coefficient at each wavenumber and constructed the quantitative absorption coefficient spectrum for the bulk solution absorption (Figure 4). Spectra from the near-surface region (z = 4.9 μm) were also converted to the absorption coefficients for comparison, representing the minimum accessible path length without introducing confinement effects on *E. coli* from mechanical compression. Characteristic *E. coli* spectral features were observed in both bulk and near-surface spectra for all trials (Figure 4b), although the quantitative absorption coefficient values tend to differ between the bulk and near-surface spectra for each trial. While the bulk absorbance was roughly consistent among all three trials, the near-surface spectra suggested differences in concentration between the trials. Overall, increased variance in the absorption coefficient was observed for the near-surface region between the three trials compared to the bulk. Possible causes for the near-surface variations in *E. coli* concentration include material transport toward the interfaces, cellular consumption of dissolved species, or adsorption to the substrate. Gravity can cause cellular settling out of the optical path, while surface properties and cell–surface interactions can promote adhesion and accumulation at the interface.[38–40] In addition to whole cells, extracellular polymeric substances can also accumulate at the surface, contributing to differences from the bulk.[4,40]

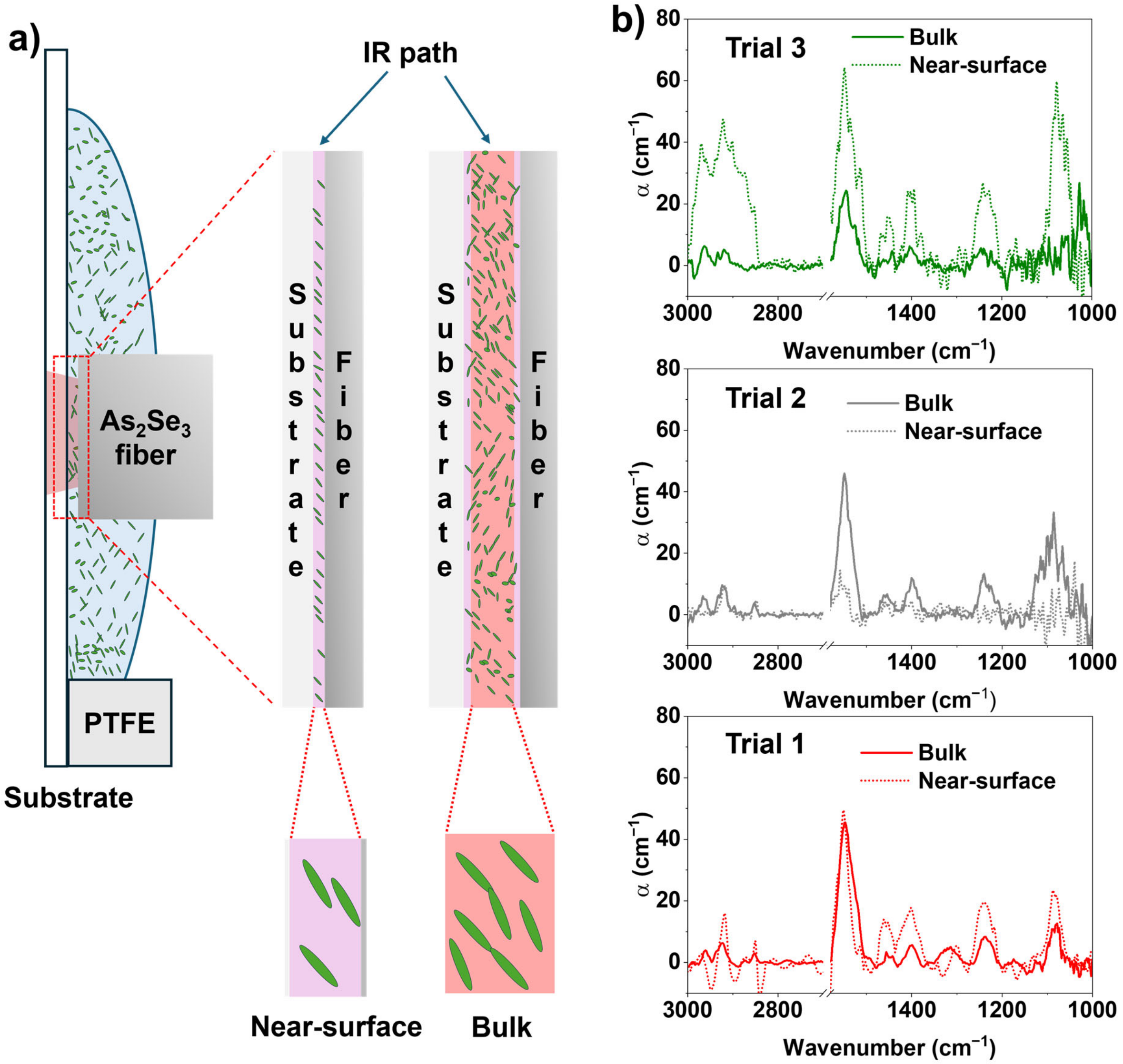

**Figure 4.** Mid-IR absorption of bulk solution vs the near-surface layer of 4.5 vol.% *E. coli* suspension. a) Schematic diagram representing bulk vs near-surface regions. b) Absorption coefficient spectra of the bulk *E. coli* solution derived from the slope of the linear regression at z ≥ 4.9 μm at each wavenumber, vs that produced only from the spectra at the near-surface region (z = 4.9 μm).

Compared to ATR-FTIR, which is heavily weighted towards the region near the internal reflection element, endo-FTIR allows detection of regions near the optical fiber and at greater distances in real time. ATR-FTIR provides information on the near-surface region, while bulk FTIR transmittance provides a summation over the entire optical path. With endo-FTIR, chemical information from both the bulk and near-surface regions can be obtained without physically separating the sample.

Endo-FTIR enables path-length-resolved measurements that are promising for many heterogeneous biological samples like cells and biofilms. Depth resolution along the optical path of the measurement could resolve contributions of concentrations of cell-associated and extracellular-matrix-associated molecules. Measurements near biomaterial interfaces could be used to investigate adsorption of proteins to substrates and cell attachment. Following cell layer or biofilm formation, changes in the spatial distribution of nutrients from cellular consumption could be measured.[4,38,41] Concentration gradients can be obtained through analysis of spectra at different path lengths.

## CONCLUSIONS

Transmission FTIR measurements of *E. coli* suspensions in water were conducted using a fabricated mid-IR fiber probe and a benchtop spectrometer over optical path lengths up to 17 µm. Measurements acquired near the substrate surface exhibited substantially greater trial-to-trial variability than bulk measurements, consistent with increased spatial heterogeneity near the interface. These results demonstrate that variable-path-length fiber transmission FTIR can distinguish spatial heterogeneities while providing optical path lengths beyond the ~1 µm surface region of conventional ATR-FTIR. The approach provides a foundation for path-length-resolved mid-IR characterization of aqueous biological samples and could be extended to many other heterogeneous systems.

## ASSOCIATED CONTENT

### Supporting Information

The supporting information is available free of charge online.

> Surface profile of fiber end before and after polishing; confirmation of optical transmission and efficacy of fiber polishing; transmittance spectra of an E. coli suspension used to estimate the optical path length; calculation of absorbance from transmittance of E. coli and DPBS; FTIR spectra of 5% aqueous BSA solution with linear fit of absorbance at 1548 $cm^{-1}$; and *E. coli* limit of detection benchmarking (pdf)

### Notes

The authors declare no competing financial interests.

## ACKNOWLEDGMENTS

We acknowledge the support from the National Science Foundation under Grant No. 2243257 (all authors). A.H. further acknowledges the support from the National Institutes of Health through the ReproEngeering T32 Training Grant Program. The experiments were performed in part in the Materials Research Laboratory at the University of Illinois. The authors acknowledge the use of facilities and instrumentation supported by NSF through the University of Illinois Materials Research Science and Engineering Center DMR-2309037.